\documentclass[pra,aps,superscriptaddress,showpacs, preprint]{revtex4-1}

\usepackage{epsf,epsfig}
\usepackage[psamsfonts]{amssymb}
\usepackage{amsmath}
\usepackage{graphicx}
\usepackage{color,soul}
\usepackage{xcolor}	

 \newcommand{\ket}[1]{|#1\rangle}
 \newcommand{\bra}[1]{\langle #1|}

 \newcommand{\e}{{\mathrm e}}

\newcommand{\Tr}{{\mathrm {Tr}}}

\newcommand{\etal}{\textit {et al.} }

 \newcommand{\half}{\frac{1}{2}}

 \newcommand{\of}{{\omega_{F}}}
 \newcommand{\oa}{{\omega_{A}}}
\newcommand{\sint}{{\sin{\theta_{n}}}}
\newcommand{\cost}{{\cos{\theta_{n}}}}

\newcommand{\rhot}{{\rho (t)}}

\begin{document}
\title{Geometric Discord Of the Jaynes-Cummings Model:
Pure Dephasing Regime}
\author{Sina Hamedani Raja\footnote{s.hamedani@sci.ui.ac.ir}}
\affiliation{Department of Physics, University of Isfahan, Isfahan, Iran}
\author{Hamidreza Mohammadi\footnote{hr.mohammadi@sci.ui.ac.ir}}
\affiliation{Department of Physics, University of Isfahan, Isfahan, Iran}
\affiliation{Quantum Optics Group, University of Isfahan, Isfahan, Iran}
\author{S. Javad Akhtarshenas\footnote{akhtarshenas@um.ac.ir}}
\affiliation{Department of Physics, University of Isfahan, Isfahan, Iran}
\affiliation{Quantum Optics Group, University of Isfahan, Isfahan, Iran}
\affiliation{Department of Physics, Ferdowsi University of Mashhad, Mashhad, Iran}

\begin{abstract}
\indent In this paper, the dynamical behaviour of the geometric discord of a system consisting of a two-level atom interacting with a quantised radiation field described by the Jaynes-Cummings model has been studied. The evolution of the system has been considered in the pure dephasing regime when the field is initially in a general pure state and the atom is initially in a mixed state. Dynamics of the geometric discord, as a measure of non-classical correlation, has been compared with the dynamics of negativity, as a measure of quantum entanglement. In particular, the influence of different parameters of system such as detuning and mixedness of the initial atomic state on the dynamics of geometric discord has been evaluated for when the field is initially in coherent and number states. It is shown that for asymptotically large times, the steady state geometric discord of the system presents a non-zero optimum value at some intermediate value of detuning.
\end{abstract}

\maketitle
\section{Introduction}
Quantum discord has proved to be a resource for performing some quantum information and computation protocols \cite{cubitt}. This approach of quantum correlation expresses a different type of correlation \cite{zurek,vedral} in comparison with entanglement, in which we are concerned about separability. Quantum discord sheds a new light on the concept of correlation in composite systems and reveals that there are some separable, i.e. disentangled,  multipartite states which possess quantum correlation and hence they can be employed as a resource for improving the quantum information processes, speeding up the quantum computation algorithms and/or performing the quantum communication protocols. Despite the fact that entanglement looks at quantum correlation from the separability point of view, quantum discord can capture quantum  correlation from the measurement perspective \cite{modi,brodutch}. However, calculation of quantum discord requires finding the best (optimised) measurement protocol to be performed on one part of the system, therefore, complicating the calculation of quantum discord and prevents one from obtaining an analytic closed-form of the formula for quantum discord in general. Indeed, quantum discord is analytically calculated only for a few families of two-qubit states \cite{lu}, for some reduced two-qubit states of pure three-qubit states, and also for a class of rank-2 mixed state of $4\otimes2$ systems \cite{gen}. On the other hand, the geometric measure of quantum discord is another measure of quantum correlation, that is in general easier to be utilised for calculations of non-classical correlations \cite{dakic}. Furthermore, Dakic \emph{et al}. have demonstrated, by using a variety of polarisation-correlated photon pairs, that non-zero quantum discord is the necessary resource for remote state preparation \cite{dakic-zeilinger}. Also, they show that the geometric measure of quantum discord is directly linked to the fidelity of remote state preparation for a broad class of states. This fact may provide an operational interpretation for this measure of quantumness.
\par
Geometric discord  is defined as the squared Hilbert-Schmidt distance between the state of the quantum system and the closest zero-discord state. For a bipartite state $\rho$ on the Hilbert space $\mathcal{H}^{A}\otimes \mathcal{H}^{B}$, geometric discord $D_{G}$ is defined as \cite{dakic}
\begin{equation}
D_{G}=\min_{\chi \in \chi_{0}} \Vert \rho -\chi \Vert ^{2},
\end{equation}
where minimisation is taken over the set of all zero-discord states $\chi_{0}$,  and $\Vert \rho -\chi \Vert ^{2}=\Tr (\rho-\chi)^{2}$ is the squared norm in the Hilbert-Schmidt space.
An exact expression of $\mathit{D}_G$ for pure $N\otimes N$ and arbitrary $2\otimes  N$ states are obtained \cite{lu1,lu2}. Consider a bipartite state $\rho$ acting on the Hilbert space $\mathbb{C}^{2}\otimes \mathbb{C}^{N}$. One can write $\rho$ in the Bloch representation as
\begin{equation}
\rho =‎\frac{1}{2N}\Big(‎‎‎\mathbb{I}^{A}‎\otimes \mathbb{I}^{B} +\sum_{i=1}^{3}‎x_{i}‎‎ ‎\sigma_{i}‎\otimes \mathbb{I}^{B}+\mathbb{I}^{A}\otimes\sum_{j=1}^{N^2-1}‎  y_{j}{‎\lambda‎}_{j}+\sum_{i=1}^{3}\sum_{j=1}^{N^2-1}t_{ij}\sigma_i‎\otimes{\lambda‎}_{j}\Big),
\end{equation}
where $\{\sigma_i\}_{i=1}^{3}$ are the usual Pauli matrices, and $\{\lambda_{j}^{}\}_{j=1}^{N^2-1}$ are  generators of $SU(N)$, fulfilling the following equations
\begin{equation}
\Tr \lambda _{i}=0, \quad \Tr(\lambda _{i}\lambda _{j})=2\delta _{ij}.
\end{equation}
Also $\vec{x}=(x_1, x_2, x_3)^t$ with $x_{i}=‎‎\mathrm{Tr}‎‎[ (‎{\sigma‎}_{i}‎\otimes \mathbb{I}^{B})\rho] \label{29}$,  and $\vec{y}=(y_1, \cdots, y_{N^2-1})^t$ with $y_{j}=‎\frac{N}{2}\mathrm{Tr}[‎(\mathbb{I}^{A}\otimes {‎\lambda‎}_{j})\rho]$  denote  local coherence vectors of  two subsystems, and $T=(t_{ij})$ with $t_{ij}=‎\frac{N}{2}‎\mathrm{Tr}[({\sigma‎}_{i}‎\otimes{\lambda‎}_{j}) \rho] \label{31}$ is the correlation matrix of the state.
 It is shown that $\mathit{D}_G$ of $ 2\otimes N$ states can be written as \cite{lu2}
\begin{equation}\label{GD formula}
D_{G}=\frac{1}{2N}\Big(\Vert \vec{x}\Vert ^{2} +\frac{2}{N}\Vert T\Vert ^{2} -\xi_{\max}\Big)=\frac{1}{2 N}(\xi_2+\xi_3) ,
\end{equation}
where  $\{\xi_{k}\}_{k=1}^{3}$ are eigenvalues of $\big(\vec{x}\vec{x}^{t}+\frac{2}{N}TT^{t}\big)$,  in non-increasing order, and $\xi_{\max}=\max\{\xi_k\}=\xi_1$. 
\par
The Jaynes-Cummings model (JCM), which describes the interaction of a single two-level atom with the quantised radiation field \cite{jaynes},  has been the subject of numerous studies in the field of quantum optics due to the fact that this system shows behaviour which clearly illustrates physical concepts such as collapse and revival. Furthermore, it has attracted much attention for the study of quantum correlations. A lot of studies have been devoted to understand the atom-field interaction and methods have been devised to create, control, and quantify quantum correlation between them both in the weak and strong coupling regimes \cite{furuichi,scheel,akhtarshenas,beaudoin,altintas,tau,jara, prakash,hessian}.
\par
In this paper, we specifically intend to investigate the quantum correlation of a system which exhibits decoherence. Decoherence is an inevitable consequence when the quantum system is interacting with a surrounding environment. If energy transfer occurs between the system and environment, the dissipation process is the most important mechanism leading to decoherence . However,  decoherence can happens even when energy is conserved in the system: in this case the process of phase damping (dephasing) reduces the quantum coherence of the system. Considering the properties of environment is important to investigate its interaction with the quantum systems and decoherence; the concepts of Markovian and non-Markovian processes that deal with the capability of the environment to possess memory of the system have been the subject of many studies. For example, the dynamical behavior of quantum correlation of bipartite systems is reviewed in \citep{franco} where the authors compared the behavior of entanglement and quantum discord in non-Markovian and Markovian processes. Moreover,  Milburn has shown that, even without explicitly model an environment and its interaction with the system, the dephasing process may occur due to quantum jumps in the evolution of the system \cite{milburn}. 
\par
Investigating the quantum discord of open systems has been the subject of several studies. For example, Jara $\etal$ have studied the discording power of an interaction between a system and its enclosing bath, both with and without the assumption of the  rotating-wave approximation (RWA) \cite{jara}. In addition, Altintas $\etal$ have used  the master equation of evolution of interacting atom and field and investigated the entanglement and quantum discord of atom-field interactions in strong coupling regime. Also, dynamics of the geometric quantum discord was studied in \cite{tau}, where the authors have employed a dissipative system of two independent atom-cavity-reservoirs, both in the strong and weak  coupling
 regimes.
 \par
In the present paper, we investigate the dynamical behavior of quantum correlations of a system consisting of a two-level atom which interacts with a single-mode quantised radiation field.  Such systems are important in operational quantum information processing. In particular, this system is suitable for performing some quantum communication tasks, because it includes a stationary qubit (atom) and a flying qudit (photon). The atom-field interaction is described by the Jaynes-Cummings model. We use the JCM in the weak coupling regime together with the RWA, and assume that our system undergoes a phase damping evolution caused by a decoherence process. The physics of this process could be quantum jumps of the atom-field state during the evolution. We focus on calculating the time-evolved state of the atom-field, where the atom is initially in a general mixed state and the field is initially in a pure   state, i.e. number and coherent states, and investigate the quantum correlations between the atom and field. Our algebraic method to calculate the geometric discord of the JCM is novel, in which the calculation of coherence vectors and correlation matrix in the JCM are carried out easily. In previous studies such as \cite{moya, shang,akhtarshenas} the pure dephasing evolution of the JCM has been considered, i.e. when the atom and field are both initially in pure states, but we consider here a more general case, when the initial state of the atom is a general mixed state. Also, our analytic results enable us to evaluate the dynamics of the quantum correlations of the JCM such as negativity and geometric discord for various values of atomic mixedness of the system, and the atom-field detuning. Our results reveal that although the asymptotic geometric discord of the system is negligible in the resonance and far resonance regions, it has a non-zero value for some intermediate detunings, and that we can find an optimum value for such asymptotic $D^{\infty}_G$ in an intermediate value of detuning. Interestingly, a similar result is mentioned in \cite{shang} for the negativity of the JCM and we show here how these two types of quantum correlation resemble each other in this view.
\par
The paper is organised as follows.
In section II, we briefly review the Jaynes-Cummings Model. Pure dephasing evolution of the system is studied in Section III. Section IV includes the calculation of the geometric discord for the two distinct initial states of the field i.e. number and coherent states. In Section V, we present our results and conclude the paper with some discussions.

\section{The Jaynes-Cummings Model}
The Jaynes-Cummings model (JCM) describes the interaction of a two-level atom with a single-mode quantised radiation field \cite{jaynes}. The Hamiltonian of this model is given by
\begin{equation}\label{1}
H=H_{0}+H_{I},
\end{equation}
where $H_{0}$ is the internal Hamiltonian of the system and $H_{I}$ describes the atom-field interaction in the rotating-wave approximation (RWA) and weak-coupling regime, where (for $\hbar=1$)
\begin{eqnarray}
H_{0}&=&\half  \omega_{A}\sigma_{z}+ \omega_{F}a^{\dagger}a,
\\
H_{I}&=& g (\sigma_{+}\otimes a + \sigma_{-}\otimes a^{\dagger}).
\end{eqnarray}
Here  $\omega_{A}$ and $\omega_{F}$ are the transition frequency of the two-level atom and the frequency of the radiation field, respectively, and $g$ denotes the atom-field coupling. Note that the RWA and weak coupling regime are valid when $g\ll \omega_{A}, \omega_{F}$. Also $\sigma_{+}$,  $\sigma_{-}=\sigma_{+}^{\dagger}$ are atomic spin-flip operators, and $\sigma_{z}$ is the atomic inversion operator which act on the atom Hilbert space $\mathcal{H}^A$. Also $a$ and $a^{\dagger}$ are annihilation and creation operators of the field acting on the field Hilbert space $\mathcal{H}^F$. The two-dimensional Hilbert space of the atom is spanned by two orthonormal states $\ket g \doteq (0,1)^{t}$ and $\ket e \doteq (1,0)^{t}$, and the field Hilbert space is spanned by the photon number states $\{\ket n =\frac{(a^{\dagger})^{n}}{\sqrt{n!}}\ket 0 \}_{n=0}^{\infty} $, where $\ket 0 $ is the vacuum state of the radiation field.
\par
To achieve the time evolution of the system, it is convenient to use the dressed-state representation of the Hamiltonian \eqref{1}. Regarding the fact that Hamiltonian \eqref{1} conserves the total number of excitations of the operator $K=(a^{\dagger}a +\half \sigma_{z})$, one can decompose the atom-field Hilbert space $\mathcal{H}=\mathcal{H}^A\otimes\mathcal{H}^F$ as $\mathcal{H}=\bigoplus_{n=0}^{\infty}\mathcal{H}_{n}$ such that $\mathcal{H}_{0}=\mathrm{span}\{\vert g,0 \rangle \}$ and $\mathcal{H}_{n+1}=\mathrm{span}\{\vert e,n \rangle ,\vert g,n+1\rangle \}$ for $n\in \{0, 1, 2,...\}$, are the eigen-subspaces of $K$ corresponding to the eigenvalues $-\frac{1}{2}$ and $(n+\frac{1}{2})$, respectively. Accordingly, Hamiltonian \eqref{1} has the eigenvalues
\begin{eqnarray}
E_{0}=-\half  \omega_{A}, \,\,\,\,\,\,\,\,\,\,\, E_{\pm}^{(n)}= \omega_{F}(n+\half )\pm  \Omega_{n},
\end{eqnarray}
with the corresponding eigenvectors
\begin{eqnarray}\nonumber \label{H-ESs}
\ket {\Phi_{0}}\,\,\,\,&=&\vert g,0 \rangle,
\\
\ket {\Phi^{(n)}_{+}}&=&\sint \ket {e,n} +\cost \ket {g,n+1},
\\ \nonumber
\ket{\Phi^{(n)}_{-}}&=&\cost \ket {e,n} -\sint \ket {g,n+1},
\end{eqnarray}
where $\tan \theta_{n}=\frac{2g\sqrt{n+1}}{-\Delta +2\Omega_{n}}$, and $\Omega_{n}=\sqrt{(\Delta/2)^{2}+g^{2}(n+1)}$ is the Rabi frequency. Also $\Delta=\oa -\of$ is the detuning parameter of the system.

\section{Pure Dephasing Evolution of the System}
Decoherence destroys the quantumness of the system and decreases the useful quantum correlations between on the parts of the system . There are several approaches to consider decoherence in a quantum system which are  responsible for quantum-classical transition. One of these approaches is based on the modified Schr\"odinger equation in such a way that the quantum coherence is automatically destroyed as the system evolves. This mechanism is called \emph{intrinsic decoherence} and has been studied in the framework of several models (see \cite{moya} and references therein). In particular, Milburn has proposed a simple modification of the standard quantum mechanics based on the assumption that for sufficiently short time steps $\gamma$ the system evolution is governed by a stochastic sequence of identical unitary transformations rather than a continuous unitary evolution \cite{milburn}. This assumption leads to a modification of Schr\"odinger equation which includes a term corresponding to the decay of quantum coherence in the energy bases. Using a Poisson model for the stochastic time steps, Milburn obtained the following dynamical master equation in the first order approximation \cite{moya, hamid, gardiner, louisell, milburn}
\begin{equation}\label{2}
\frac{d}{dt}\rhot =-i[H,\rhot]-\frac{\gamma}{2}[H,[H,\rhot]],
\end{equation}
where $\rhot$ is the density matrix of the atom-field in any given time $t\geq 0$.
Equation \eqref{2} has the following formal solution \cite{moya,xu}
\begin{equation}\label{3}
\rhot =\sum _{k=0}^{\infty}\frac{(\gamma t)^{k}}{k!}M^{k}(t)\rho(0)M^{\dagger^{k}}(t),
\end{equation}
such that $\rho(0)$ is the initial state of the system and
\begin{equation}
M^{k}(t)=H^{k}\exp(-iHt)\exp(-\frac{\gamma t}{2}H^{2}).
\end{equation}
\par
In order to find the time-evolved state $\rhot$, we  need to expand the initial state $\rho(0)$ in terms of the dressed-states, i.e.  the Hamiltonian eigenbases \eqref{H-ESs}, as
\begin{eqnarray}\label{initial state}
\rho(0) &=&
\ket {\Phi_{0}}\bra {\Phi_{0}} \rho(0) \ket {\Phi_{0}}\bra {\Phi_{0}}
+\sum_{\alpha=\pm}\sum_{n=0}^{\infty}\ket {\Phi_{\alpha}^{(n)}}\bra {\Phi_{\alpha}^{(n)}} \rho(0) \ket {\Phi_{0}}\bra {\Phi_{0}}  \\ \nonumber
&+&\sum_{\beta =\pm}\sum_{m=0}^{\infty}\ket {\Phi_{0}}\bra {\Phi_{0}} \rho(0) \ket {\Phi_{\beta}^{(m)}}\bra {\Phi_{\beta}^{(m)}}
+ \sum_{\alpha,\beta =\pm}\sum_{m,n=0}^{\infty}
\ket {\Phi_{\alpha}^{(n)}}\bra {\Phi_{\alpha}^{(n)}} \rho(0) \ket {\Phi_{\beta}^{(m)}}\bra {\Phi_{\beta}^{(m)}}. \nonumber
\end{eqnarray}
Inserting this in Eq. \eqref{3} we get
\begin{eqnarray}\label{Rho-t1}
\rhot &=&\ket {\Phi_{0}}\bra {\Phi_{0}}\bra {\Phi_{0}} \rho(0) \ket {\Phi_{0}} \\
&+&\sum_{\alpha =\pm}\sum_{n=0}^{\infty}\ket {\Phi_{\alpha}^{(n)}}\bra {\Phi_{0}}\exp \Big( -i (\omega_{\alpha}^{n})t-\frac{\gamma t}{2}(\omega_{\alpha}^{n})^{2} \Big) \bra {\Phi_{\alpha}^{(n)}} \rho(0) \ket {\Phi_{0}}\nonumber
\\
&+&\sum_{\beta =\pm}\sum_{m=0}^{\infty}\ket {\Phi_{0}}\bra {\Phi_{\beta}^{(m)}}\exp \Big( i (\omega_{\beta}^{m})t-\frac{\gamma t}{2}(\omega_{\beta}^{m})^{2} \Big) \bra {\Phi_{0}} \rho(0) \ket {\Phi_{\beta}^{(m)}}\nonumber \\
&+&\sum_{\alpha,\beta =\pm}\sum_{m,n=0}^{\infty}\ket {\Phi_{\alpha}^{(n)}}\bra {\Phi_{\beta}^{(m)}}\exp \Big( -i (\omega_{\alpha\beta}^{nm})t-\frac{\gamma t}{2}(\omega_{\alpha\beta}^{nm})^{2} \Big) \bra {\Phi_{\alpha}^{(n)}} \rho(0) \ket {\Phi_{\beta}^{(m)}},\nonumber
\end{eqnarray}
where we have defined
\begin{equation}
\omega_{\alpha\beta}^{nm}=E^{(n)}_{\alpha}-E^{(m)}_{\beta}, \qquad \omega_{\alpha}^{(n)}=E^{(n)}_{\alpha}-E_0.
\end{equation}
\par
Now let us suppose that, initially at $t=0$, the system is found in the product state
\begin{equation}\label{RhoA-RhoF}
\rho(0)=\rho^{A}(0)\otimes \rho^{F}(0),
\end{equation}
such that $\rho^{A}(0)$ is the initial state of the atom and is considered to be a  mixed state
\begin{equation}\label{RhoA}
\rho^{A}(0)=p\ket e \bra e + (1-p) \ket g \bra g, \qquad 0 \le p \le 1,
\end{equation}
and $\rho^{F}(0)$, the initial state of the field,  is assumed to be a  pure state
\begin{equation}\label{RhoF}
 \rho^{F}(0)=\ket \eta\bra\eta,
\end{equation}
with $\ket \eta =\sum_{n=0}^{\infty}b_{n}\ket n$, where the complex coefficients $b_{n}s$ satisfy the normalisation condition $\sum_{n=0}^{\infty}\vert b_{n}\vert^{2}=1$. In section \ref{GD section}, we will fix  the coefficients $b_n$ for  two special cases, namely  number states and coherent states.
Accordingly, inserting Eqs. \eqref{RhoA-RhoF}-\eqref{RhoF} into Eq. \eqref{Rho-t1} and using  orthonormal bases $\{\ket{e_1}\equiv\ket e , \ket{e_2}\equiv\ket g \}$ for the atomic Hilbert space, one can find, after tedious but straightforward calculations, the following representation for the density matrix $\rho(t)$
\begin{equation}\label{Rho-t2}
\rhot=
\left( \begin{array}{c|c}
\hat{A}(t) & \hat{C}(t) \\ \hline
\hat{C}^{\dagger}(t) & \hat{B}(t)
\end{array} \right),
\end{equation}
where $\hat{A}(t)$, $\hat{B}(t)$, and $\hat{C}(t)$,  operators acting on the field Hilbert space $\mathcal{H}^F$, are defined by
\begin{align}
\hat{A}(t)&=p\hat{A}^{(e)}(t)+(1-p)\hat{A}^{(g)}(t),
\\
\hat{B}(t)&=p\hat{B}^{(e)}(t)+(1-p)\hat{B}^{(g)}(t),
\\
\hat{C}(t)&=p\hat{C}^{(e)}(t)+(1-p)\hat{C}^{(g)}(t),
\end{align}
where matrix elements of operators $\hat{A}^{(e,g)}(t)$, $\hat{B}^{(e,g)}(t)$, and $\hat{C}^{(e,g)}(t)$ in the Fock bases $\{\ket n\}^{\infty}_{n=0}$  are given by
\begin{align}\label{115}
A_{nm}^{(e)}(t)&=b_{n}b_{m}^{‎\star‎}\Bigg [‎\sin^{2}‎(‎\theta_{n})\Big[‎‎\sin^{2}‎(‎\theta‎_{m})‎\exp‎\big(-i\omega^{nm}_{++}t-‎\frac{‎\gamma‎ t}{2}‎(\omega^{nm}_{++})^{2}\big)\nonumber
\\
&+\cos^{2}‎ (\theta _{m})\exp‎\big(-i\omega^{nm}_{+-}t-‎\frac{‎\gamma‎ t}{2}‎(\omega^{nm}_{+-})^{2}\big)\Big]\nonumber
\\
&+\cos^{2}(‎\theta‎_{n})\Big[‎\sin^{2}‎ (‎\theta‎_{m})‎\exp‎\big(-i\omega^{nm}_{-+}t-‎\frac{‎\gamma‎ t}{2}‎(\omega^{nm}_{-+})^{2}\big)\nonumber
\\
&+‎\cos^{2}‎ (‎\theta‎_{m})‎\exp‎\big(-i\omega^{nm}_{--}t-‎\frac{‎\gamma‎ t}{2}‎(\omega^{nm}_{--})^{2}\big)\Big]\Bigg],
\end{align}
\begin{align}
A_{nm}^{(g)}(t)&=‎\frac{1}{4}b_{n+1}b_{m+1}^{‎\star‎}‎\sin‎( 2\theta_{n})\sin‎(‎ 2\theta‎_{m})\Big[‎\exp‎\big(-i\omega^{nm}_{++}t-‎\frac{‎\gamma‎ t}{2}‎(\omega^{nm}_{++})^{2}\big)\nonumber
\\
&-\exp‎\big(-i\omega^{nm}_{+-}t-‎\frac{‎\gamma‎ t}{2}‎(\omega^{nm}_{+-})^{2}\big)\nonumber
\\
&-‎‎\exp‎\big(-i\omega^{nm}_{-+}t-‎\frac{‎\gamma‎ t}{2}‎(\omega^{nm}_{-+})^{2}\big)\nonumber
\\
&+\exp‎\big(-i\omega^{nm}_{--}t-‎\frac{‎\gamma‎ t}{2}‎(\omega^{nm}_{--})^{2}\big)\Big],
\end{align}
\begin{align}
B_{nm}^{(e)}(t)&=‎\frac{1}{4}b_{n-1}b_{m-1}^{‎\star‎}‎\sin‎( 2\theta_{n-1})\sin‎(‎ 2\theta‎_{m-1})\Big[‎\exp‎\big(-i\omega^{n-1,m-1}_{++}t-‎\frac{‎\gamma‎ t}{2}‎(\omega^{n-1,m-1}_{++})^{2}\big)\nonumber
\\
&-\exp‎\big(-i\omega^{n-1,m-1}_{+-}t-‎\frac{‎\gamma‎ t}{2}‎(\omega^{n-1,m-1}_{+-})^{2}\big)\nonumber
\\
&-‎‎\exp‎\big(-i\omega^{n-1,m-1}_{-+}t-‎\frac{‎\gamma‎ t}{2}‎(\omega^{n-1,m-1}_{-+})^{2}\big)\nonumber
\\
&+\exp‎\big(-i\omega^{n-1,m-1}_{--}t-‎\frac{‎\gamma‎ t}{2}‎(\omega^{n-1,m-1}_{--})^{2}\big)\Big],
\end{align}
\begin{align}\label{120}
B_{nm}^{(g)}(t)&=b_{n}b_{m}^{‎\star‎}\Bigg[‎\cos‎^{2}(\theta _{n-1})\Big[\cos^{2}(\theta _{m-1})‎\exp‎\big(-i\omega^{n-1,m-1}_{++}t-‎\frac{‎\gamma‎ t}{2}‎(\omega^{n-1,m-1}_{++})^{2}\big)\nonumber
\\
&+‎\sin‎^{2}(\theta _{m-1})\exp‎\big(-i\omega^{n-1,m-1}_{+-}t-‎\frac{‎\gamma‎ t}{2}‎(\omega^{n-1,m-1}_{+-})^{2}\big) \Big]\nonumber
\\
&+\sin^{2}(\theta _{n-1})\Big[\cos^{2}(\theta_{m-1})‎‎\exp‎\big(-i\omega^{n-1,m-1}_{-+}t-‎\frac{‎\gamma‎ t}{2}‎(\omega^{n-1,m-1}_{-+})^{2}\big)\nonumber
\\
&+\sin^{2}(\theta_{m-1})\exp‎\big(-i\omega^{n-1,m-1}_{--}t-‎\frac{‎\gamma‎ t}{2}‎(\omega^{n-1,m-1}_{--})^{2}\big)\Big] \Bigg],
\end{align}
\begin{align}
C_{nm}^{(e)}(t)&=‎‎\frac{1}{2}b_{n}b_{m-1}^{‎\star‎}‎‎\sin‎( 2‎\theta‎_{m-1})\Bigg [‎\sin^{2}‎(‎\theta_{n})\Big[‎\exp‎\big(-i\omega^{n,m-1}_{++}t-‎\frac{‎\gamma‎ t}{2}‎(\omega^{n,m-1}_{++})^{2}\big)\nonumber
\\
&-\exp‎\big(-i\omega^{n,m-1}_{+-}t-‎\frac{‎\gamma‎ t}{2}‎(\omega^{n,m-1}_{+-})^{2}\big)\Big]\nonumber
\\
&+‎\cos^{2}(‎\theta‎_{n})\Big[\exp‎\big(-i\omega^{n,m-1}_{-+}t-‎\frac{‎\gamma‎ t}{2}‎(\omega^{n,m-1}_{-+})^{2}\big)\nonumber
\\
&-\exp‎\big(-i\omega^{n,m-1}_{--}t-‎\frac{‎\gamma‎ t}{2}‎(\omega^{n,m-1}_{--})^{2}\big)\Big]\Bigg],
\end{align}
\begin{align}
C_{nm}^{(g)}(t)&=\frac{1}{2}b_{n+1}‎b_{m}^{‎\star‎}‎\sin‎‎(2\theta _{n})\Bigg[\cos^{2}(\theta _{m-1})‎‎\Big[‎\exp‎\big(-i\omega^{n,m-1}_{++}t-‎\frac{‎\gamma‎ t}{2}‎(\omega^{n,m-1}_{++})^{2}\big)\nonumber
\\
&-‎\exp‎\big(-i\omega^{n,m-1}_{-+}t-‎\frac{‎\gamma‎ t}{2}‎(\omega^{n,m-1}_{-+})^{2}\big)\Big]\nonumber
\\
&+\sin^{2}(\theta_{m-1})‎‎\Big[\exp‎\big(-i\omega^{n,m-1}_{+-}t-‎\frac{‎\gamma‎ t}{2}‎(\omega^{n,m-1}_{+-})^{2}\big)\nonumber
\\
&+\exp‎\big(-i\omega^{n,m-1}_{--}t-‎\frac{‎\gamma‎ t}{2}‎(\omega^{n,m-1}_{--})^{2}\big)\Big]\Bigg].
\end{align}

\section{geometric discord of the System} \label{GD section}
In this section, we turn  our attention on the geometric quantum discord of the state given by Eq. \eqref{Rho-t2}.
To do so,  we should first note that the state \eqref{Rho-t2} is supported, actually,  on a $2\otimes \infty$ Hilbert space and investigating the correlation properties of this system may appear to an impossible task. However, as we will show in the following sections, by considering some appropriate initial states for the field, our state at a given time $t> 0$ can be represented  by a  $2 N \times 2 N$ matrix  for a finite $N$.
In continue, we express two examples of these preparations.
\subsection{Number state as initial state of the field }
We first consider the case that the field is  prepared, initially, in a given number state $\ket k $, i.e.  the coefficients $b_{n}$  are set to be $b_{n}=\delta_{nk}$. In this particular case, the  atom-field state $\rho(t)$ of Eq. \eqref{Rho-t2} is supported on a $2‎\otimes 3‎$  Hilbert space. By setting $\{\ket{k-1},\ket{k},\ket{k+1}\}$ as the orthonormal bases of the field, one can represent the corresponding density matrix as
\begin{equation}\label{Rho-t3}
\rho_{k}(t)=\left(
\begin{array}{cccccc}
(1-p)B_{k-1}(t)&0&0&0&-(1-p)C_{k-1}(t)&0\\
0&pA_{k}(t)&0&0&0&pC_{k}(t)\\
0&0&0&0&0&0\\
0&0&0&0&0&0\\
-(1-p)C^{\ast}_{k-1}&0&0&0&(1-p)\big(A_{k-1}(t)+‎\delta_{k,0}\big)&0\\
0&pC^{\ast}_{k}(t)‎&0&0&0&pB_{k}(t)
\end{array}\right),
\end{equation}
where
\begin{align}
&A_{k}(t)=‎\frac{1}{4}\bigg(‎2+\frac{‎\Delta‎^{2}}{2‎\Omega^{2}_{k}‎}‎+\bigg(2-‎\frac{‎\Delta‎^{2}}{2‎\Omega^{2}_{k}‎}‎\bigg)‎\cos‎(2‎\Omega_{k}t)‎\exp‎(-2‎\gamma t ‎‎\Omega^{2}_{k})\bigg) \label{38},
\\
&C_{k}(t)=\frac{g‎\sqrt{k+1}‎}{4‎\Omega _{k}‎}‎\bigg(‎\frac{‎\Delta‎}{‎\Omega‎_{k}}‎(1-‎\cos‎(2‎\Omega_{k}t)‎\exp‎(-2‎\gamma t ‎‎\Omega^{2}_{k}))+2i‎‎\sin‎‎(2‎\Omega_{k}t)‎\exp‎(-2‎\gamma t ‎‎\Omega^{2}_{k})\bigg)  \label{39},
\\
&B_{k}(t)=‎\frac{g^{2}(k+1)}{2‎\Omega ^{2}_{k}‎}‎\bigg(1-‎\cos‎(2‎\Omega_{k}t)‎\exp‎(-2‎\gamma t ‎‎\Omega^{2}_{k})\bigg)  \label{40}.
\end{align}
 This density matrix can be used to obtain the coherence vector $\vec{x}$ and correlation matrix $T$, and hence the geometric discord.

\subsection{Coherent state as initial state of the field}
As the  second example, we consider the case that the field is initially in a coherent state $\ket{\alpha}$, i.e. $b_n=\e^{-\half \vert \alpha \vert ^{2}}\frac{\alpha^{n}}{\sqrt{n!}}$.
Although this initial condition implies a $2\otimes \infty$ support for  $\rho(t)$, the Poissonian distribution of photon numbers in coherent states allows us to truncate the dimension of the field Hilbert space to a finite one. Numerically, it is seen that if we set $|\alpha|=\sqrt{5}$ the coefficients $b_{n}$s will be negligible for $n>30$; Indeed, the ratio $\frac{\vert b_{n+1} \vert^{2}}{\vert b_{n}\vert ^{2}}$ becomes less than $10^{-10}$ for large enough values of $n$, so it is reasonable to restrict the dimension of the field Hilbert space to an appropriate $N$. To fulfill the adequate accuracy in the case of $|\alpha|=\sqrt{5}$, numerical calculations have shown that $N=30$ is sufficient, such that $\Tr[\rho (t)]\cong 1$, so we truncate $\mathcal{H}^F$ to a 30-dimensional space. This approach was also employed in \cite{akhtarshenas} to calculate the negativity of the JCM where the system undergoes a unitary evolution, and the consideration of the initial states of the atom and field are the same as the present paper. We use the same approach to evaluate the dephasing evolution of the JCM, and it can be seen that just by setting the dephasing parameter $\gamma$ equal to zero, i.e. the unitary evolution,  the results of \cite{akhtarshenas} for the state of the system can be achieved. Nevertheless, in the case of dephasing evolution, the 25-dimensional field Hilbert space which is used in \cite{akhtarshenas} is no longer accurate in numerical calculations, so we have expanded this dimension to thirty. 
\par
Now we focus on calculating the $\mathit{D}_G$ of the atom-field state while the field is initially in the mentioned coherent state. Consequently, each block of the density matrix \eqref{Rho-t2} will be a $30 \times 30 $ matrix acting on the field. If we let $\{{\lambda}_{j}\}_{j=1}^{30^{2}-1}$ to be the generators of $SU(30)$ and use the Pauli matrices as the generators of the $SU(2)$, and utilise the representation in \eqref{Rho-t2}, we have
\begin{align}
x_{1}(t)&=\Tr_{F} \Big[\hat{C}^{\dagger}(t)+\hat{C}(t)\Big],
\\
x_{2}(t)&=\Tr_{F} \Big[-i\hat{C}^{\dagger}(t)+i\hat{C}(t)\Big],
\\
x_{3}(t)&=\Tr_{F} \Big[\hat{A}(t)-\hat{B}(t)\Big],
\end{align}
and
\begin{align}
t_{1j}(t)&=15\Tr_{F} \Big[\hat{\lambda}_{j}\big(\hat{C}^{\dagger}(t)+\hat{C}(t)\big)\Big],
\\
t_{2j}(t)&=15\Tr_{F} \Big[\hat{\lambda}_{j}\big(-i\hat{C}^{\dagger}(t)+i\hat{C}(t)\big)\Big],
\\
t_{3j}(t)&=15\Tr_{F} \Big[\hat{\lambda}_{j}\big(\hat{A}(t)-\hat{B}(t)\big)\Big],
\end{align}
where $j=1,\dots,30^{2}-1$ and $\Tr_{F}$ is the partial trace over the field which can be readily applied on the matrix representation of the operators $\hat{A}(t)$, $\hat{B}(t)$, $\hat{C}(t)$ and $\hat{C}^{\dagger}(t)$. Having the coherence vector $\vec{x}$ and correlation matrix $T$ in hand, one can calculate the geometric discord via Eq. \eqref{GD formula}. Meanwhile, an explicit representation of the generators of $SU(30)$ is required; The method proposed in \cite{georgi} has expressed a set of straightforward equations to achieve both Cartan sub-algebra and non-diagonal generators of $SU(N)$.

\section{Results and Discussions}

Based on the analytic calculations of the pure dephasing evolution of the atom-field system expressed in the previous sections,  we have calculated the geometric discord $\mathit{D}_G$ of the state for different values of the parameters $g$, $\Delta$, $\gamma$ and $p$. The evolution of the state of the system and hence the dynamics of the quantum correlations strongly depend on the initial state of the system. Therefore, in  all of the following calculations we suppose that the atom can be found, initially, in a mixture of the ground and excited states,  and the field is in a pure initial state. For the pure state of the field we consider two distinct cases: the number state and the coherent state. We mostly focus our attention on the quantum correlations of  the JCM in pure dephasing evolution, but it is worthwhile to evaluate our results for $\gamma=0$, i.e. the unitary evolution, and then compare them with the results of the situation when the dephasing process is present.

\subsection{Initiating the field in a number state}
We first consider the case that the field is initially prepared in a number state, so that the behaviour of the system over time can be readily extracted by examining Eq.  \eqref{Rho-t3}.
\begin{itemize}
\item{ {\it Unitary Evolution  ($\gamma=0$)}\\
Suppose that   the initial state of the field is set to be vacuum state $\ket 0$.  Figure  \ref{Fig1}-(a)  shows the time behaviour of $\mathit{D}_G$ when the atom is initially in the excited state $\ket e$, and for some regular values of detuning $\Delta$, where RWA is valid.
\begin{figure}[ht!]
\centering
‎\includegraphics[width=18cm]{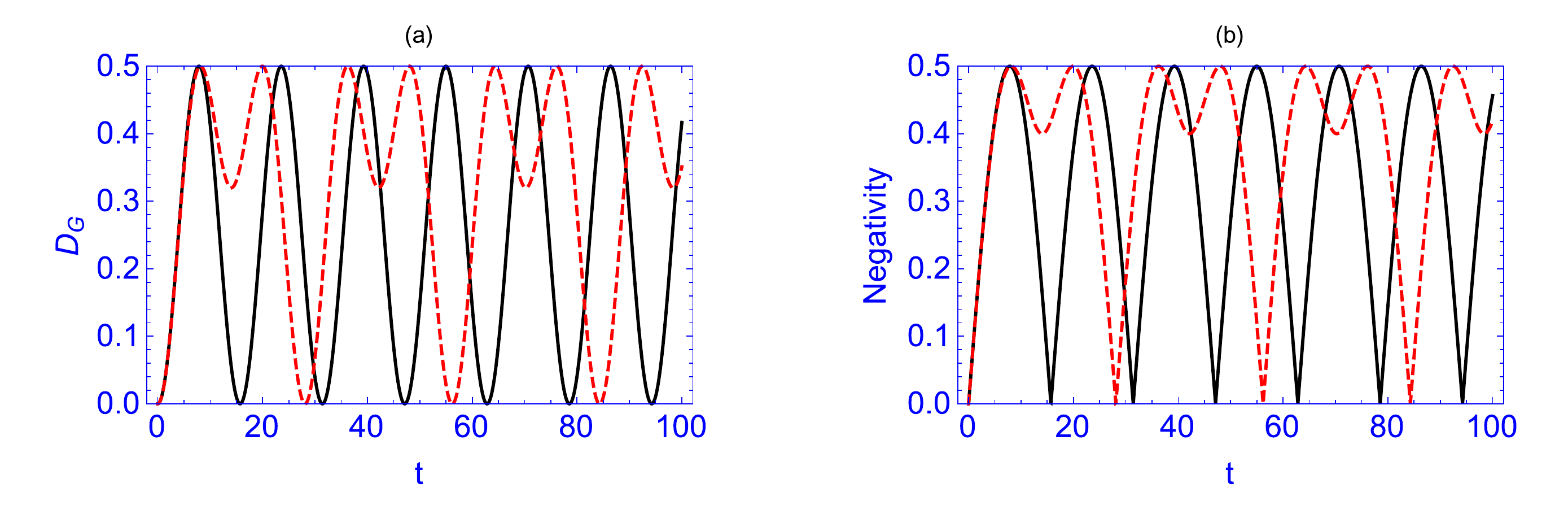}
\caption{(Color online) Time evolution of  $\mathit{D}_G$ and negativity  when the field is initially in  the vacuum state $\ket 0$, and  $\gamma=0$ (unitary evolution), $p=1$, $g/\omega_A=0.1$ with $\Delta=0$ [solid black line] and  $\Delta=0.1$ [dashed red line].}‎
\label{Fig1}
\end{figure}
Since  evolution is unitary, the pure initial state of the system will remain pure during the evolution. As it is mentioned in Ref. \cite{luo}, for a $m\otimes m$ pure state $\ket{\Psi}=\sum_{i=1}^m\sqrt{s_i}\ket{i}\ket{i}$, the geometric discord is related to the generalised concurrence  as
\begin{equation}
D_G(\Psi)=1-\sum_{i=1}^{m}s_i^2=\frac{1}{2}C^2(\Psi),
\end{equation}
where  $C(\Psi)$ is the generalised concurrence of $\ket{\Psi}$ \cite{rungta}.  Figure \ref{Fig1}-(b) shows the dynamics of the negativity of the JCM with the same assumptions. As it is clear, the time behaviour of $\mathit{D}_G$ and negativity are the same up to a scale factor. We can see that as the collapse and revival occurs, the atom and the field becomes periodically correlated and de-correlated.}
\item{ {\it Dephasing regime ($\gamma \neq0$)}\\
When the dephasing parameter $\gamma$ is not zero, the story is different. Decoherence suppresses the coherence oscillations (collapse and revival) of the evolution and after sufficient time, it leads the system to get a stationary state. In Figs. \ref{Fig2} and \ref{Fig3} we have plotted  $D_{G}$ versus time for different values of $p$ and $\Delta$ when the initial state of the field is $\ket 1$.
\begin{figure}[ht!]
\centering
‎\includegraphics[width=18cm]{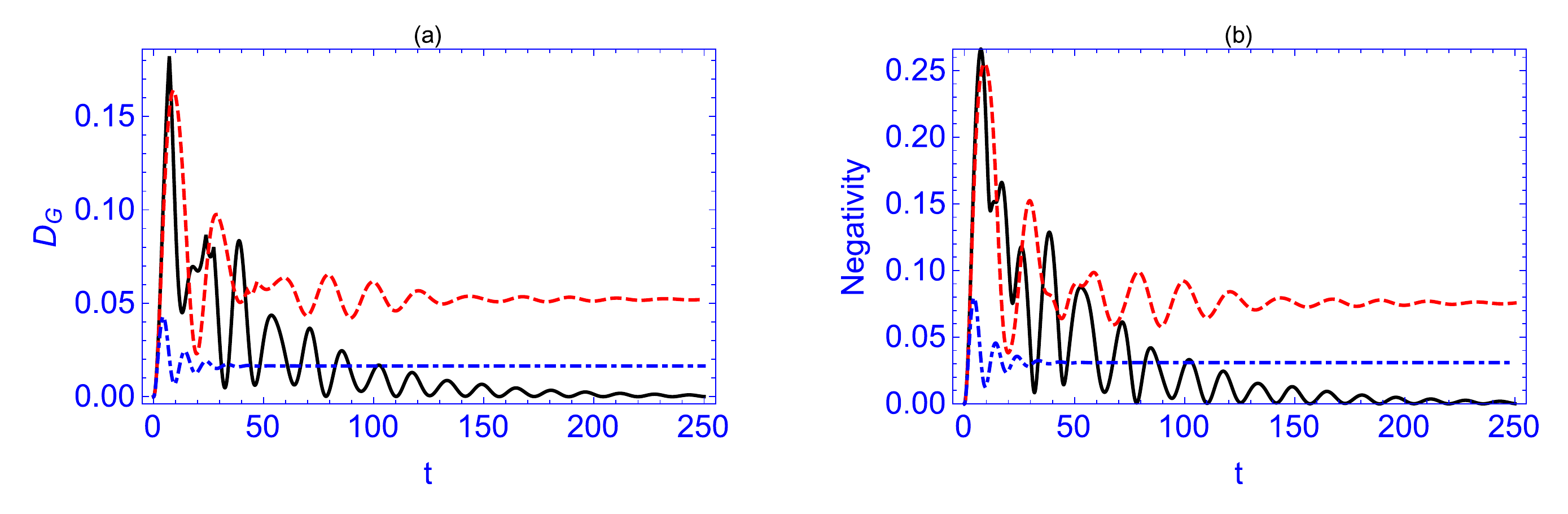}
\caption{(Color online) Time evolution of  $\mathit{D}_G$ and negativity when field is initially in $\ket 1$ for $g/{\omega_A}=0.1$, $\gamma=0.5$ and $p=0.5$ with $\Delta=0$ [solid black line], $\Delta=0.2$ [dashed red line] and $\Delta=0.6$ [dot-dashed blue line].}
\label{Fig2}
\end{figure}
\begin{figure}[ht!]
\centering
‎\includegraphics[width=18cm]{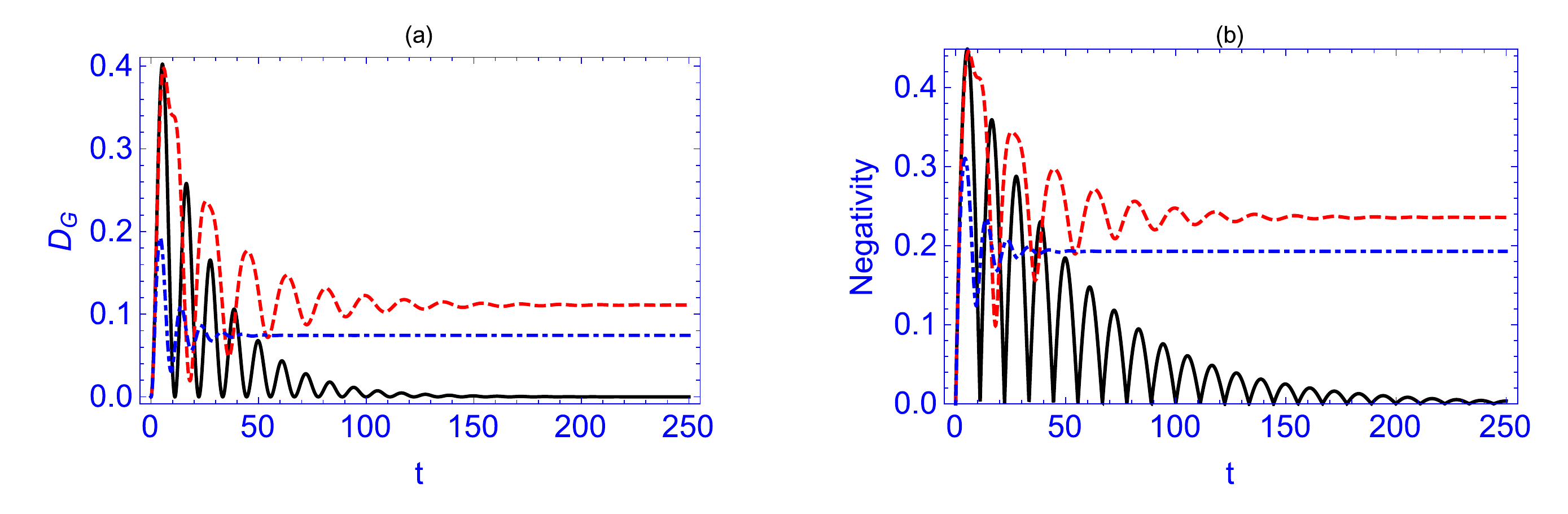}
\caption{(Color online) Time evolution of $\mathit{D}_G$ and negativity when field is initially in $\ket 1$ for $g/\omega_A=0.1$, $\gamma=0.5$ and $p=1$ with $\Delta=0$ [solid black line], $\Delta=0.2$ [dashed red line] and $\Delta=0.6$ [dot-dashed blue line].}
\label{Fig3}
\end{figure}
Since elements of the density matrix are proportional to $\e ^{-\gamma t}$, they vanish after times $t\gg \gamma$,  for $\gamma\neq0$. So the system reaches a stationary state, asymptotically. Hence we have the following time-independent elements for the density matrix of the system at the asymptotically large times:
\begin{align}
&\bar{A}_{k}(\Delta,k,g)=‎\frac{1}{4}\bigg(‎2+\frac{‎\Delta‎^{2}}{(\Delta^{2}/2)+2g^2(k+1)‎}\bigg),\nonumber
\\
&\bar{C}_{k}(\Delta,k,g)=\frac{g‎\Delta‎\sqrt{k+1}‎}{\Delta^{2}+4g^2(k+1)‎},
\\
&\bar{B}_{k}(\Delta,k,g)=‎\frac{g^2(k+1)}{(\Delta^{2}/2)+2g^2(k+1)‎}.\nonumber
\end{align}
As it is clear,  the detuning parameter $\Delta$ have an important role in the asymptotic geometric discord $D_{G}^{\infty}$. For instance, for $\Delta=0$ we have $\bar{A}_{k}(0)=\half$, $\bar{C}_{k,k-1}(0)=0$ and $\bar{B}_{k}(0)=\half$ i.e
the density matrix \eqref{Rho-t3} becomes diagonal and consequently $\mathit{D}_G^\infty=0$. So it seems that we need a nonzero $\Delta$ in order to have nonzero $\mathit{D}_G^\infty$. Likewise, the evolution of the system barely creates any correlation for large values of detuning. Accordingly,  the asymptotic geometric discord $\mathit{D}_G^\infty$ may have a maximum value at a finite optimum value of detuning, $\Delta_{opt.}$. Figure \ref{Fig4} expresses the asymptotic geometric discord $\mathit{D}_G^\infty$ versus $\Delta$ when the field is initially in $\ket 1 $ and for different values of $p$. Finally, Figure \ref{Fig5} summarise the previous results in a 3D plot of $\mathit{D}_G^\infty$ versus $\Delta$ and $p$.
\begin{figure}[th!]
\centering
‎\includegraphics[width=10cm]{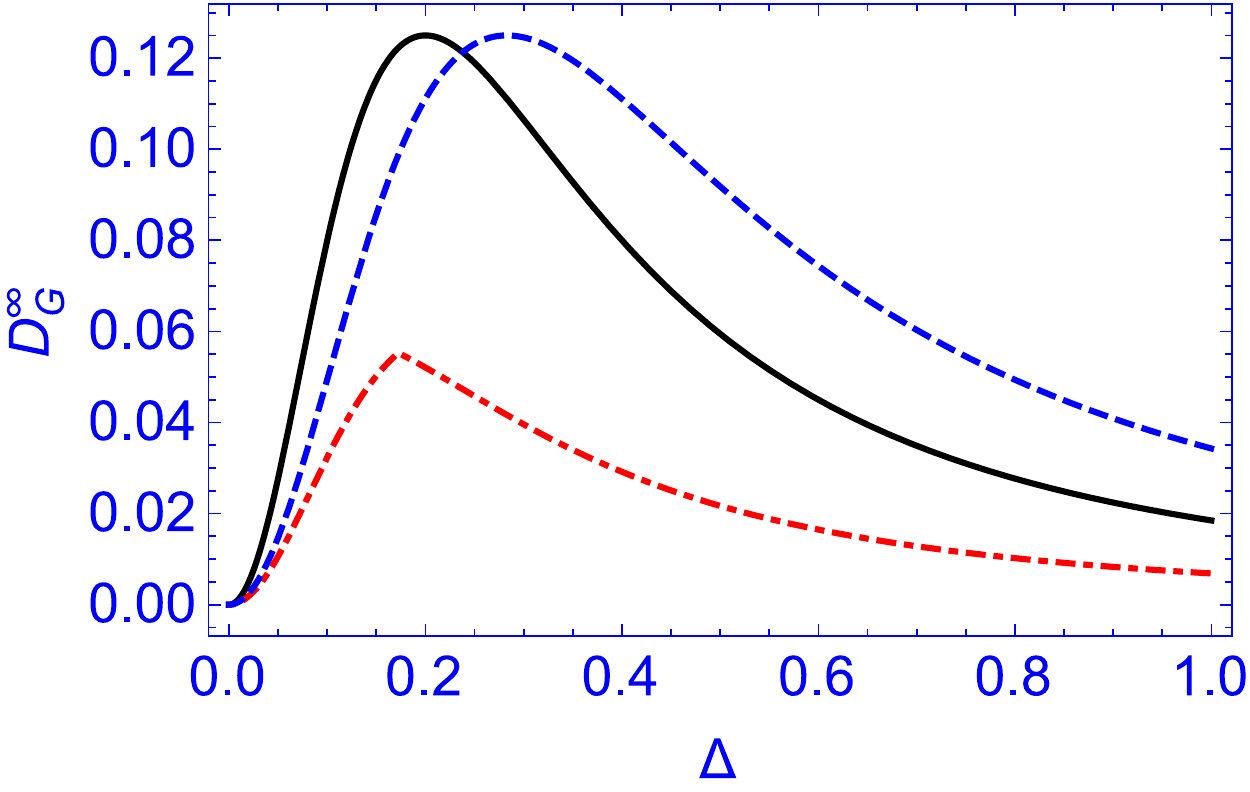}
\caption{(Color online)  The asymptotic geometric discord $\mathit{D}_G^\infty$ as a function of $\Delta$ when the field is initially in $\ket 1$ and $g/\omega_A=0.1$ with $p=0$ [solid black line], $p=0.5$ [dot-dashed red line] and  $p=1$ [dashed blue line].}
\label{Fig4}
\end{figure}
These figures reveal that the value of $D_{G}^\infty$ seems to be not symmetric with respect to the parameter $p$. Since the one-dimensional dressed state  $\ket {g,0}$ does not change through the evolution, it turns out that the rules of $p$ and $(1-p)$ differ in evolution, leading therefore to the above mentioned asymmetry.
\begin{figure}[ht!]
\centering
‎\includegraphics[width=12cm]{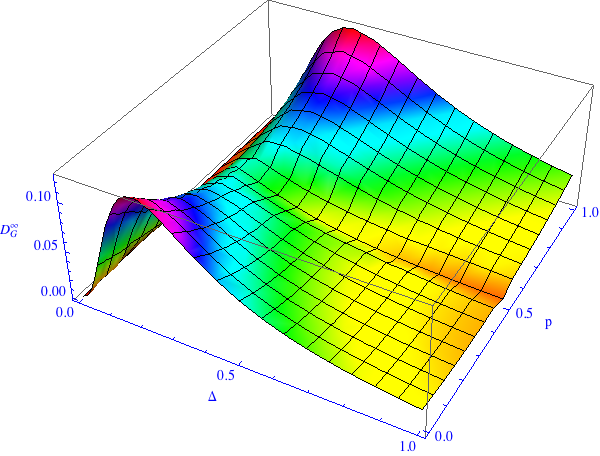}
\caption{(Color online) The asymptotic geometric discord $\mathit{D}_G^\infty$ as a function of $\Delta$ and $p$ when the field is initially in the number state $\ket 1$.}
\label{Fig5}
\end{figure}}

\end{itemize}

\subsection{Initiating the field in a coherent state}
We now check our results in the case where the field is prepared initially in a coherent state $\ket{\alpha}$. In the following case, we fix $|\alpha|=\sqrt{5}$, so it is sufficient to use only the 30-dimensional truncated subspace of the field Fock space.
\begin{itemize}
\item{ {\it Unitary evolution ($\gamma=0$)}\\
First, suppose the evolution is unitary, i.e. $\gamma=0$.  Figure \ref{Fig6} shows  $\mathit{D}_G$ of the JCM when the atom is initially in the excited state , i.e. $p=1$,  and in resonance with the field, i.e. $\Delta=0$. Authors of \cite{akhtarshenas} have calculated the negativity and mutual information of the JCM with the similar assumptions. Evidently, apart from a difference in scaling, Fig. \ref{Fig6} exhibits coincidence of the results: the dynamical behavior of $\mathit{D}_G$ coincides with the dynamical behavior of negativity presented in \cite{akhtarshenas}.
\begin{figure}[ht!]
\centering
‎\includegraphics[width=12cm]{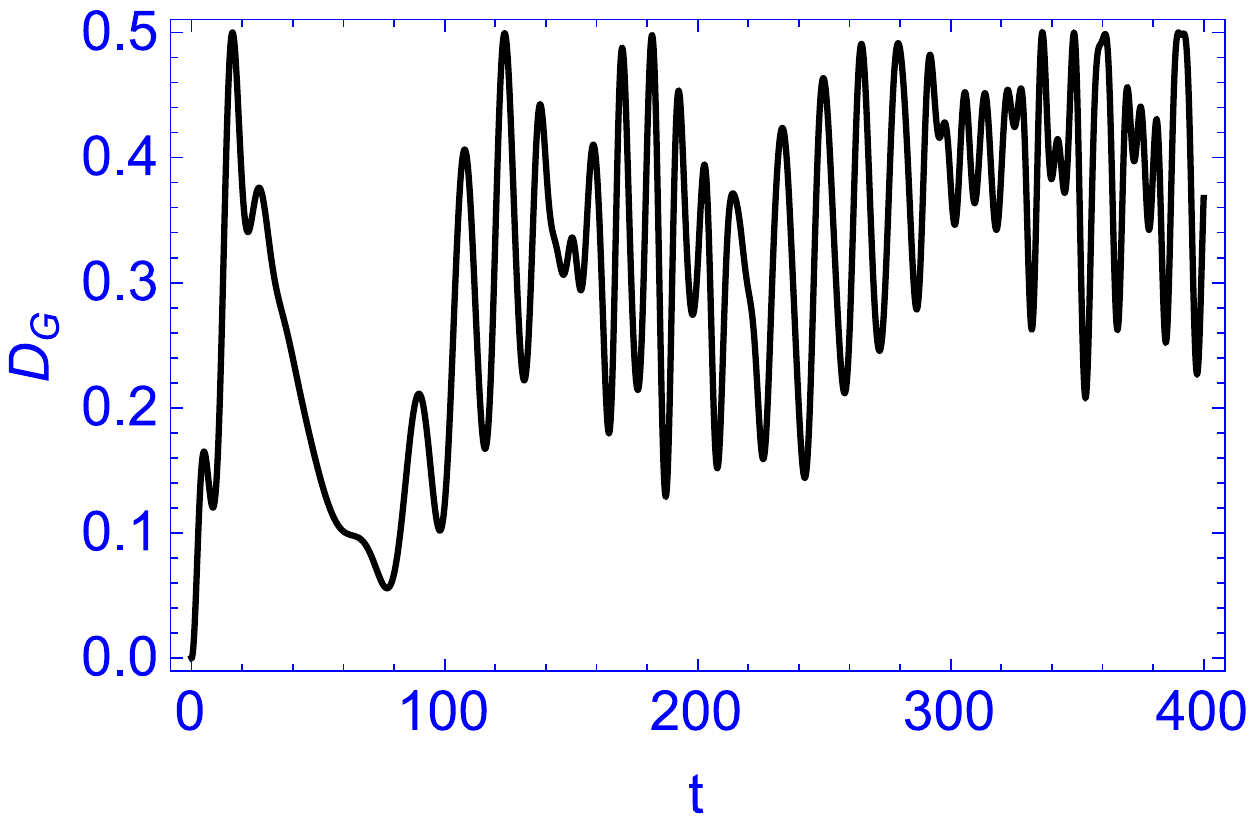}
\caption{(Color online) Time evolution of $\mathit{D}_G$ when field is initially in $\ket{\alpha}$ for $\gamma=0$ (unitary evolution), $g/\omega_A=0.1$, $\Delta=0$ and $p=1$ with $|\alpha|=\sqrt{5}$.}
\label{Fig6}
\end{figure}}
\item{ {\it Dephasing regime ($\gamma \neq 0$)}\\
The dephasing mechanism tends to demolish the coherent elements of the density matrix Eq.\eqref{Rho-t2} (suppose $b_n=\e^{-\half \vert \alpha \vert ^{2}}\frac{\alpha^{n}}{\sqrt{n!}}$ for $n\leq 30$). It is evident that for the asymptotically large times ($t\gg \gamma$), the coherent term of the density matrix, $ \hat{C}(t)$, approaches zero unless in the cases where $‎(‎E^{(n)}_{\pm}-E^{(m)}_{\pm})= 0$; This happens only for eight elements, namely four diagonal and four non-diagonal elements. For large times, however, it is not difficult to see that under the resonant condition $\Delta=0$,  only the diagonal elements remain and consequently the density matrix becomes zero-discord. In the opposite situation, when the detuning $\Delta$ is large enough, the non-diagonal elements  vanish for the asymptotically large times. Similar to the conclusion in the previous case, there exists an optimum value of detuning which gives the maximum  value for the geometric discord of the stationary state. Figure \ref{Fig7} shows the asymptotic geometric discord $\mathit{D}_G^\infty$ versus $\Delta$ when the initial state of the field is in the coherent state $\ket{\alpha}$ with $|\alpha|=\sqrt{5}$.
\begin{figure}[ht!]
\centering
‎\includegraphics[width=12cm]{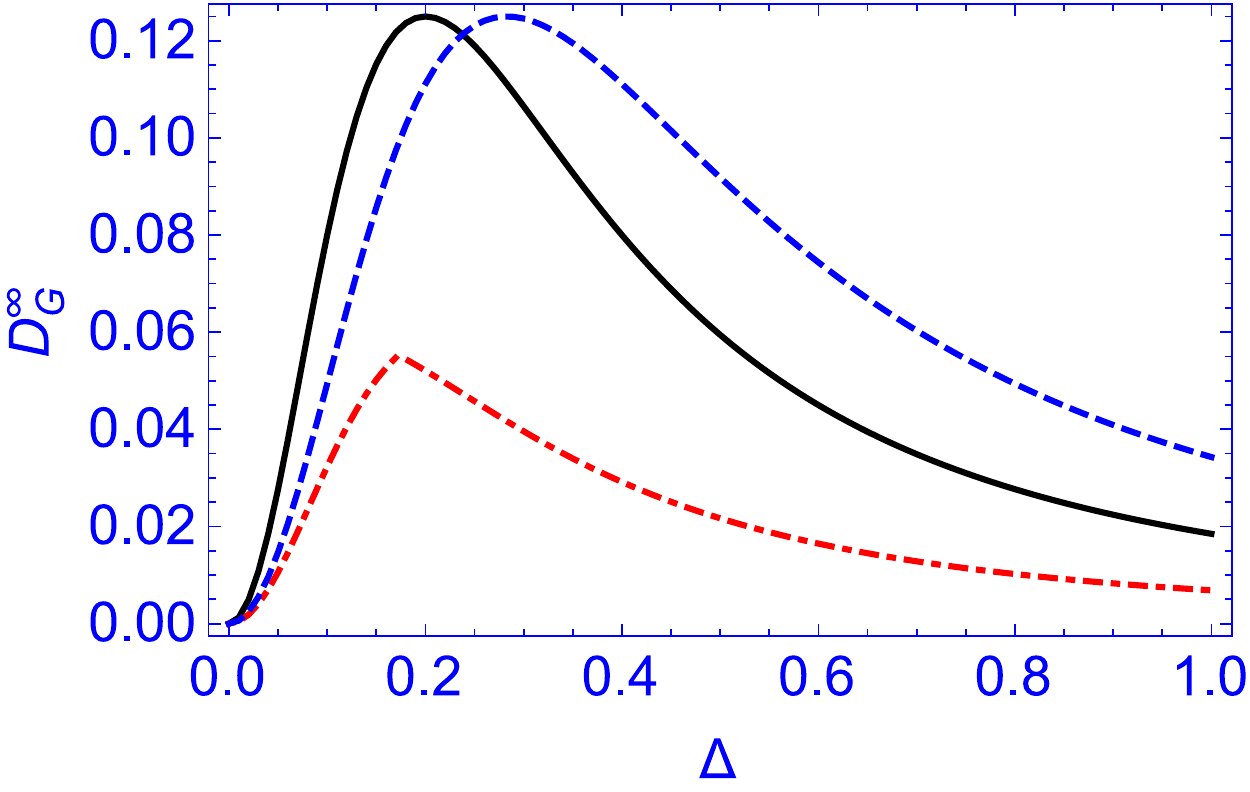}
\caption{(Color online)  $\mathit{D}^{\infty}_G$ as a function of $\Delta$ when field is initially in the coherent state $\ket{\alpha}$, with $p=0$ [solid black line],  $p=0.5$ [dot-dashed red line] and $p=1$ [dashed blue line].}
\label{Fig7}
\end{figure}}
\end{itemize}
\par
To discuss some aspects of our results, our calculations reveals that for the asymptotic case, i.e. when $t \longrightarrow \infty$, the support of the density operator of the system lies only on a 4-dimensional subspace, irrespective of the initial state. One can vary this support by changing the system parameters,  for instance the detuning $\Delta$. In this case, the asymptotic geometric discord depends on $\Delta$, $g$ and $p$ (also $k$ for number states). For instance, for the case of $p=0.5$, the supports of the stationary states coincide for the coherent and the number initial states of the field. This fact is obvious from Figures \ref{Fig4} and \ref{Fig7}; these figures are exactly the same. For other values of $p$, the rank of the support remains unchanged but the above mentioned coincidence does not occur.\\
Also, it is worthwhile to discuss how correlation can survive in an off-resonant system while they do not survive in the resonant case. The quantum correlation can resulted either from a resonant  transfer of energy between the atom and the field, or from dispersive energy shifts affecting the atoms and the photons when they are not resonant \cite{raimond,brune,schleich}.  The fact is that, although the system gains significant amounts of correlation in the resonant case, this correlation is vulnerable to dephasing and can not survive. On the other hand, the quantum correlation of the system in off-resonant case is rather robust to the dephasing process and are longer-lasting. Strictly speaking, the energy transfer process dominates the resonant atom and field, while the latter dominates the far off-resonant system. 
Our results indicate that although the contributions of the two above processes change as we alter the detuning, the ability of the system to gain correlation is generally remarkable for detunings between zero up to some intermediate value (notice Figure \ref{Fig1}), while this ability decreases for larger detunings; So we can conclude that the ability of the system to gain correlation does not immediately wane by an increase in detuning, but the reduction of this ability occurs when the detuning is greater than some intermediate value. In addition, the immunity of the system to the dephasing is very important, and it strongly depends on the initial state of the system, and also on the amount of detuning. Speaking about the immunity, the coherent elements (off-diagonal elements) of expression shown in \eqref{Rho-t1} are fragile to dephasing, and vanish at large times, so their influence in maintaining correlation of the system is highly dependent on the dephasing process. But since the energy of the atom-field system is conserved, the population elements of \eqref{Rho-t1} are invariant and their correspondence in the quantum correlation of the system can survive. We speculate that for the initial state \eqref{initial state}, the resonant transfer of energy between atom and field which dominates the behaviour of the correlation in a resonant system is highly relevant to the coherent elements of \eqref{Rho-t1}, while for an off-resonant system, the amounts of correlation is more relevant to the population elements, and accordingly, is more robust against the dephasing processes. The fact is that detuning changes the energy levels of the atom-field system and the coherent and population elements of the evolution \eqref{Rho-t1}, so it influences both the ability of system to create correlation and the immunity of correlation to the dephasing. Mathematically, consider the density matrix of \eqref{Rho-t1} to be the sum of two matrices as below
\begin{equation}
\rho(t)=M ^{E}+\bar{M}(t)
\end{equation}
where the $M ^{E}$ shows the diagonal elements of $\rho(t)$ corresponding with energy levels of Hamiltonian (5) and $\bar{M}(t)$ corresponds with off-diagonal elements of the density matrix (14). Expanded in detail, we have
\begin{equation}
M ^{E}=\sum_{\alpha=\mp}\sum_{n=0}^{\infty}\ket {\Phi_{\alpha}^{(n)}}\bra {\Phi_{\alpha}^{(n)}}\bra {\Phi_{\alpha}^{(n)}} \rho(0) \ket {\Phi_{\alpha}^{(n)}}+\ket {\Phi_{0}}\bra {\Phi_{0}}\bra {\Phi_{0}} \rho(0) \ket {\Phi_{0}},
\end{equation}
and since the energy of the system is conserved, this matrix is time-invariant, whereas $\bar{M}(t)$ vanishes at large times because of dephasing. Accordingly, we have $\underset{t\to \infty }{\lim\rho(t)} =M ^{E}$. It is evident that the off-diagonal elements of $M ^{E}$ in the standard atom-field representation (not dressed states) are crucial in quantum correlation of the system, and they depend on both the initial state of the system, $\rho(0)$, and the parameter $\theta_{n}(\Delta,g)$ (see \eqref{H-ESs}) which determines the energy levels of the system. So for a given $\rho(0)$, the parameter $\theta_{n}(\Delta,g)$ governs $M^{E}$ and consequently determines the amounts of correlation of the system at asymptotically large times. Therefore, both the parameters $\Delta$ and $g$ should be taken into account to study the asymptotic correlation of the system (also $n$ is important when the initial state of the field is a number state). With respect to our assumption for the initial state \eqref{initial state}, the state $\rho^\infty=M^{E}$ is correlation-free when the system is in resonance, whereas, it contains correlations for the off-resonant case. Since asymptotic geometric discord is a positive function vanishing at $\Delta=0$ and large values of $\Delta$, according to the Rolle's theorem it must have a maximum at some intermediate detuning i.e. $\Delta_{op}$. 
\par
In summary, we have investigated the dynamical behavior of the geometric discord and negativity of a bipartite system including a two-level atom, prepared initially in a mixture of ground and excited states, and a quantised radiation field, prepared in a general pure number/coherent initial state. The internal interaction is modelled by the JCM in the weak coupling approximation. The evolution of the density matrix of the composite system is studied under pure dephasing process. The results of this paper show that dephasing, induced by intrinsic decoherence, is competing with internal interaction to create an asymptotic value of quantum correlation after some coherence oscillation. Hence the amount of quantum correlation of the system reaches a stationary value, asymptotically. The dynamical and asymptotic behavior of the geometric discord depend on the initial conditions and the system parameters. Indeed, the effects of dephasing can be amplified or weakened by adjusting both the parameters of the model and the initial conditions. We have also shown that the asymptotic geometric discord reaches its maximum value at an intermediate non-zero value of atom-field detuning. This system can be employed for implementation of the quantum remote state preparation protocol and the results help to improve the fidelity of this task.

\section*{Acknowledgments}
The authors wish to thank The Office of Graduate
Studies of The University of Isfahan for their support.


\end{document}